\documentstyle[12pt,epsfig]{article}
\begin{document}
\title{Nonequilibrium opinion spreading on\\ 2D small-world networks}
\author{Juli\'an Candia\\
{\small\it Center for Complex Network Research and Department of Physics,}\\ 
{\small\it University of Notre Dame, Notre Dame, IN 46556, USA}\\
{\small E-mail: jcandia@nd.edu}}
\maketitle

\begin{abstract}
Irreversible opinion spreading phenomena are studied on small-world networks 
generated from 2D regular lattices by means of the 
magnetic Eden model, a nonequilibrium kinetic model for the growth of binary mixtures in contact 
with a thermal bath. 
In this model, the opinion or decision of an individual 
is affected by those of their acquaintances, but opinion changes 
(analogous to spin flips in an Ising-like model) are not allowed. 
Particularly, we focus on aspects inherent to the underlying 2D nature of the 
substrate, such as domain growth and cluster size distributions. 
Larger shortcut fractions are observed to favor long-range 
ordering connections between distant clusters across the network, 
while the temperature is shown to drive the system across an order-disorder transition, 
in agreement with previous investigations on related equilibrium spin systems. 
Furthermore, the extrapolated phase diagram, as well as the correlation length critical exponent,
are determined by means of standard finite-size scaling procedures. 
\end{abstract}

\section{Introduction}
Nowadays, statistical physics is providing valuable tools and insight into 
several emerging, steadily growing interdisciplinary fields of science \cite{wei91,wei00,oli99,sta06}.  
In particular, many efforts have focused on the mathematical modeling 
of a rich variety of social phenomena, such as social influence and self-organization, cooperation, 
opinion formation and spreading, evolution of social structures, etc 
(see e.g. \cite{sch71,wei71,gal82,gal90,gal97,szn00,kup02,ale02,kle03,ben05,gon06,smi06,can06,des06,bor07,can07}).

Four decades ago, seminal observations by Milgram \cite{mil67} showed that the mean social 
distance between a pair of randomly selected individuals was astonishingly short, 
typically of a few degrees. Since then, the phenomenon of average shortest path-lengths that scale logarithmically 
with the system size, known as {\it small-world effect}, has ubiquitously been found in a huge diversity 
of different real networks, such as the Internet and the World Wide Web, ecological and food webs, 
power grids and electronic circuits, genome and metabolic reactions, 
collaboration among scientists and among Hollywood actors, and many others.
Furthermore, empirical observations also showed that local neighborhoods are generally 
highly interconnected \cite{wat99,alb02,new02,dor03,new06}. 

The well-studied classical random graphs, which are networks built by linking nodes at random, 
display the small-world effect but have much lower connectivities than usually observed in real networks.   
In this context, the {\it small-world networks} were proposed few years ago \cite{wat99, wat98} 
as a realization of complex networks having short mean path-lengths 
(and hence showing the small-world effect) as well as large connectivities. 
Starting from a regular lattice, a small-world network is built by randomly adding or 
rewiring a fraction $p$ of the initial number of links. Even a small fraction of added or rewired links 
provides the shortcuts needed to produce the small-world effect, thus displaying a global behavior close to 
that of a random graph, 
while preserving locally the ordered, highly connected structure of a regular lattice. 
It has been shown that this small-world regime is reached for any given disorder probability $p > 0$, 
provided only that the system size $N$ is large enough (i.e. $N>N_c$, 
where the critical system size is $N_c\propto 1/p$) \cite{new99a, bar00}.

Small-world networks, showing the appropriate topological features observed in real social networks, can thus be  
meaningfully used as substrates in order to investigate different social phenomena. 
Indeed, this motivated the study of spin models defined on small-world networks, in which 
spin states denote different opinions or preferences \cite{ale02,can06,bar00,git00,sve02,her02,med03}. 
Under this interpretation, the  
coupling constant describes the convincing power between interacting individuals, 
which is in competition with the ``free will'' given by the thermal noise \cite{ale02}, while  
a magnetic field could be used to add a bias that can be interpreted as ``prejudice'' or ``stubbornness'' \cite{sve02}.   

Very recently, irreversible opinion spreading phenomena have been studied in 1D small-world networks by means 
of the so-called magnetic Eden model (MEM) \cite{can06}, 
a nonequilibrium kinetic growth model in which the deposited particles have an intrinsic spin and grow 
in contact with a thermal bath \cite{van94,can01}. 
According to the growth rules of the MEM, which are given in 
the next Section, the opinion or decision of an individual would be affected by those of their 
acquaintances, but opinion changes (analogous to spin flips in the Ising model) would not occur. 
The MEM defined on a 1D small-world network presents a second-order phase transition taking place 
at a finite critical temperature for any value of the rewiring probability $p>0$, a phenomenon analogous to 
observations reported previously in the investigation of equilibrium spin systems \cite{can06}.

Within the context of these recent developments, the aim of this work is to study the behavior of the MEM 
growing on small-world networks generated from 2D regular lattices. Particular emphasis is put on aspects 
inherent to the underlying 2D nature of the substrate, such as domain growth and cluster size distributions. 
Moreover, the critical behavior associated to the observed thermally-driven order-disorder phase transitions 
is also studied and discussed in detail. 
Notice that, although this work is mainly motivated by social phenomena, a magnetic language is 
adopted throughout. As commented above, physical concepts such as temperature and magnetization, 
spin growth and clustering, ferromagnetic-paramagnetic phase transitions, etc, 
can be meaningfully re-interpreted in sociological/sociophysical contexts. 

This paper is organized as follows: 
in Section 2, details on the model definition and the simulation method are given; 
Section 3 is devoted to the presentation and discussion of the results, and Section 4 contains the conclusions. 

\section{The Monte Carlo simulation method}

In this work, we consider the 2D, nearest-neighbor, adding-type 
small-world network model \cite{bol88,new99b}. Starting with a 2D lattice of $N$ 
sites and $2N$ bonds, a network realization is built by adding new links connecting pairs of 
randomly chosen sites.  
For each bond in the original lattice, a shortcut is added with probability $p$. 
During this process, multiple connections between any pair of sites are avoided, as well as  
connections of a site to itself. Since the original lattice bonds are not rewired, 
the resulting network remains always connected in a single component.  
On average, $2pN$ shortcuts are added and the mean coordination number is $\langle z\rangle =4(1+p)$. 
Note that $p$ can also be regarded as the mean shortcut fraction relative to the number of 
fixed lattice bonds.    

Once the network is created, a randomly chosen up or down spin (the {\it seed}) is deposited on 
a random site. Then, the growth takes place by adding, one by one, further spins to the immediate neighborhood 
(the perimeter) of the growing cluster, taking into account the corresponding interaction energies. By analogy
to the Ising model, the energy $E$ of a configuration of spins is given by
\begin{equation}
E = - \frac{J}{2} \sum_
{\langle ij\rangle} S_iS_j ,
\label{energy}   
\end{equation}
where $S_i= \pm 1$ indicates the orientation of the spin for each occupied site (labeled by the 
subindex $i$), $J>0$ is the ferromagnetic 
coupling constant between nearest-neighbor (NN) spins, and 
$\langle ij\rangle$ indicates that the summation is taken over all pairs of occupied NN sites.
As with other spin systems defined on complex networks, the magnetic interaction between any pair 
of spins is only present when a network link (either a lattice bond or a shortcut) connects their sites. 

Setting the Boltzmann constant equal to unity ($k_{B} \equiv 1$), 
the probability for a new spin to be added to the (already grown) cluster is
defined as proportional to the Boltzmann factor exp$(-\Delta E /T)$, 
where $\Delta E$ is the total energy change involved and $T$ is the absolute temperature of the thermal bath. 
Energy and temperature are measured in units of the NN coupling constant, $J$, throughout. 
At each step, all perimeter sites have to be considered and the probabilities of adding a new (either up 
or down) spin to each site must be evaluated. 
Using the Monte Carlo simulation method, all growth probabilities are first computed and normalized, and then    
the growing site and the orientation of the new spin are both determined by means of a pseudo-random number.
Although the configuration energy of a MEM cluster, given by Eq.(\ref{energy}), resembles the Ising Hamiltonian, it 
should be noticed that the MEM is a nonequilibrium model in which new spins are 
continuously added, while older spins remain frozen and are not 
allowed to flip. 
The growth process naturally stops after the deposition of $N$ particles, 
when the network becomes completely filled. 

Notice that, following the original Eden model \cite{ede58} and earlier work on the MEM \cite{can06,van94}, 
the growth process is initiated by a single seed, whose location and orientation is randomly chosen. 
Different Monte Carlo runs use each time a different, randomly chosen seed. Using many seeds simultaneously 
may affect the subsequent dynamics, although the critical features of the system are not expected to depend on the
initial number of seeds, as follows from the invariance of interface growth exponents in the Eden model 
under different seed geometries \cite{jul85,fre85,hir86}.   

For any given set of defining parameters (i.e. the network size $N$, the shortcut-adding  
probability $p$ and the temperature $T$), ensemble averages were calculated 
over $10^2$ to $10^3$ different (randomly generated) networks, and considering 
typically $10$ to $10^2$ runs with different (randomly chosen) seeds for each network configuration. 
Since all normalized growth probabilities have to be recalculated at each deposition step, 
the resulting update algorithm is rather slow. Involving a considerable computational 
effort, this work presents extensive Monte Carlo simulations that cover the whole 
shortcut-adding probability range $0\leq p\leq 1$
for different network sizes up to $N=10^6$.  

\section{Results and discussion}

In order to examine the effect of shortcuts in the MEM growth process, let us first discuss 
some snapshot configurations at a qualitative level. Figure 1 shows typical samples at different  
growth stages and for different temperatures, for a network of size $N=10^4$. The small-world network 
we consider here has one single shortcut that links two opposite sites across a diagonal of the lattice, 
namely the nodes located at rectangular coordinates $(25,75)$ and $(75,25)$, where the coordinate origin 
$(0,0)$ is taken at the lower left corner of each plot. 
On the latter, we place an up spin as the starting seed. Up spins are shown as black dots, down spins 
as red dots, while white dots indicate empty nodes.   
The upper (lower) panel corresponds to configurations in which the system has grown to 
the $30\%$ ($90\%$) of the total network size.  

\begin{figure}[t]
\centerline{{\epsfxsize=6.truein\epsfysize=4.truein\epsffile{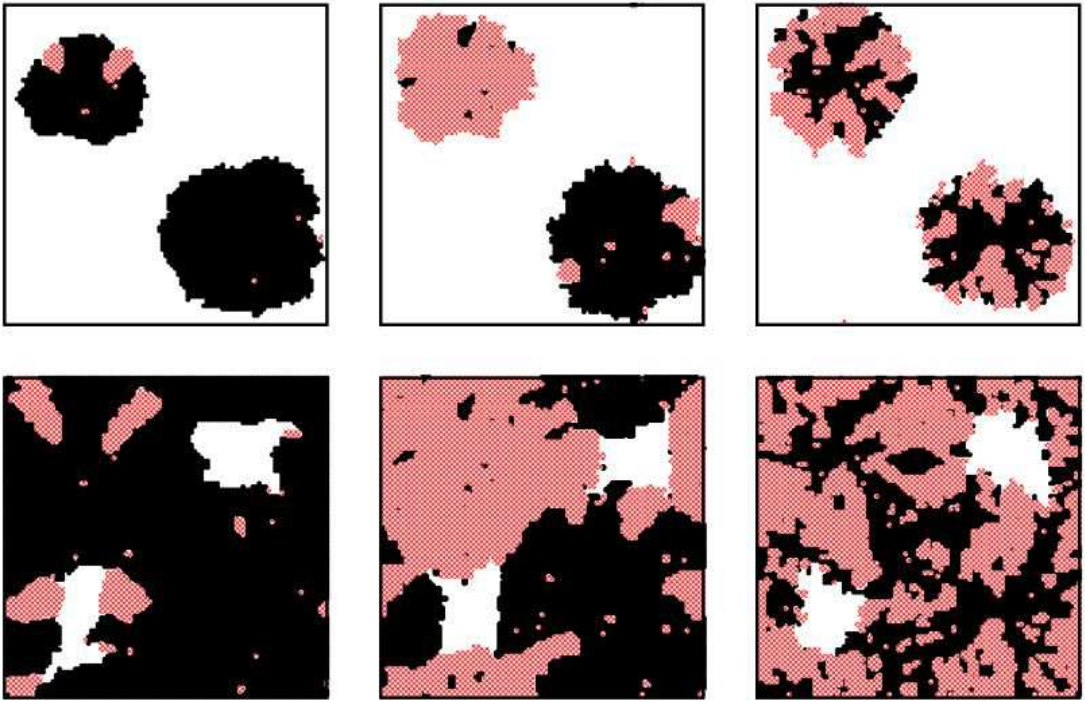}}}
\caption{Snapshots of MEM clusters at different growth stages: $30\%$ (upper panel) and 
$90\%$ (lower panel) of the total network size ($N=10^4$). 
Different configurations of up spins (black), down spins (red), and 
empty nodes (white) correspond to different temperatures: $T=0.5$ (left), $T=0.6$ (center), and $T=0.9$ (right).
One single shortcut links the nodes located at coordinates $(25,75)$ and $(75,25)$ across a diagonal of the lattice, 
where the coordinate origin $(0,0)$ is taken at the lower left corner of each plot.}
\label{fig1}
\end{figure}

The upper configurations show clearly the effect of the presence of shortcuts. 
Even one single shortcut, as we consider here, is capable of affecting substantially the growth of the 
system, creating two distinct domains of similar size. For low temperatures (upper left snapshot), 
the domains grow ordered and the orientation of the seed prevails. Increasing the temperature (upper center snapshot), 
changes of magnetization are more likely, eventually leading to the formation of two large, oppositely oriented domains. 
Notice also the presence of many smaller islands within each large domain. For even larger temperatures  
(upper right snapshot), domains become disordered due to the large thermal noise. 
As particle deposition proceeds further, the two large domains grow and eventually merge together, as shown by 
the lower panel configurations. Note, in particular, the lower center snapshot showing the clash of oppositely 
oriented domain interfaces.   

\begin{figure}[t]
\centerline{{\epsfxsize=3.8in \epsfysize=2.8in \epsfbox{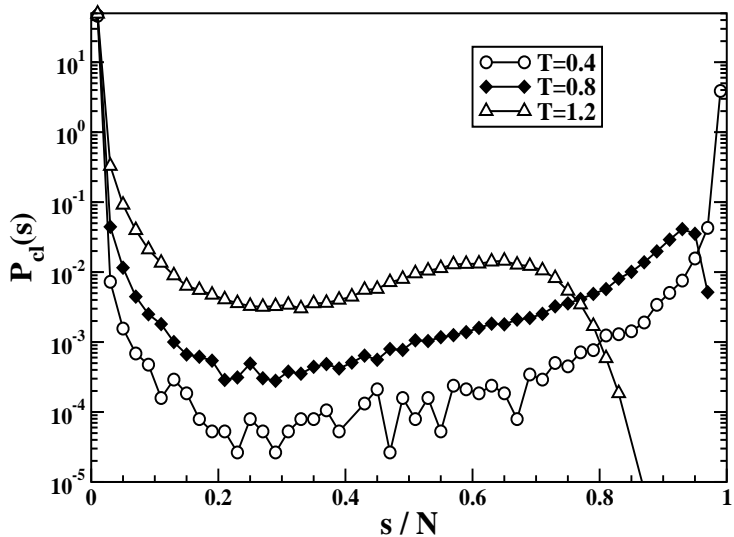}}}
\caption{Cluster size probability distribution for a system of size $N=10^4$, shortcut-adding probability $p=0.5$, and 
different temperatures, as indicated.}
\label{fig2}
\end{figure}  

The snapshot configurations of Figure 1 show rich domain structures that are highly dependent on both the temperature 
and the density of shortcuts. A quantitative description of the observed phenomena can be obtained by determining 
the cluster size probability distribution, the size of the largest cluster, and the number of different 
clusters occurring at different temperatures and shortcut densities, as follows. 

Once the system is completely grown, one can identify, on the final configuration, 
all different connected clusters of spins with the same orientation. By calculating 
the histogram of cluster sizes (and averaging over different seeds, Monte Carlo runs, and shortcut configurations), one can 
determine $P_{cl}(s)$, the probability of occurrence of a cluster of size $s$. This is shown in Figure 2 for small-world 
networks with shortcut-adding probability $p=0.5$, size $N=10^4$, and temperatures $T=0.4, 0.8,$ and $1.2$, as indicated. 
Due to thermal 
fluctuations, all distributions show an absolute maximum at small cluster sizes ($s/N\ll 1$). However, the probability 
distribution for large clusters reveals intrinsic differences in the growth mode, which depend on the temperature. 
Apart from statistical fluctuations, the distributions of clusters of size $s/N\geq 0.2$ grow monotonically with $s$. 
However, for $T=0.4$, the monotonic trend continues up to $s/N=1$, where the distribution has a local maximum, 
while, for larger temperatures, the distributions show abrupt cutoffs at $s/N<1$, the more evident the larger the temperature. 

\begin{figure}[t]
\centerline{{\epsfxsize=3.8in \epsfysize=2.8in \epsfbox{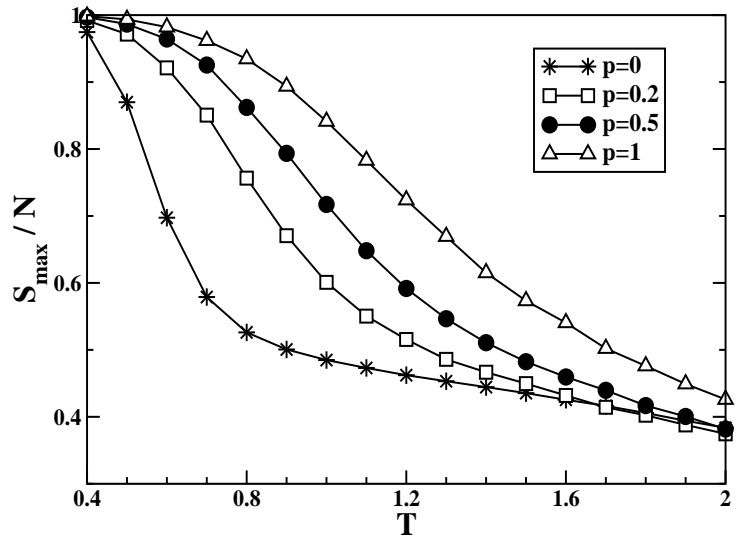}}}
\caption{Mean largest cluster size as a function of temperature, for a system of size $N=10^4$ 
and different values of the shortcut-adding probability, as indicated.}
\label{fig3}
\end{figure}

Figure 3 displays the mean largest cluster size, $S_{max}/N$, as a function of the temperature, for a system of size $N=10^4$ 
and different values of the shortcut-adding probability: $p=0, 0.2, 0.5,$ and $1$. All plots are 
monotonically decreasing with $T$, 
as expected from the onset of thermal noise. Notice the merging of the plots in the high-temperature end, 
which is due to the fact that magnetic interaction effects tend to vanish at high temperatures, 
thus leading to an essentially random deposition of up and down spins.  
Considering the effect of increasing $p$, one observes a corresponding 
increase in the size of the largest cluster, that agrees with the general observation of shortcuts increasing the 
degree of order of a spin system defined on small-world network structures. Indeed, larger shortcut fractions favor long-range 
ordering connections between distant clusters across the network.

\begin{figure}[t]
\centerline{{\epsfxsize=3.8in \epsfysize=2.8in \epsfbox{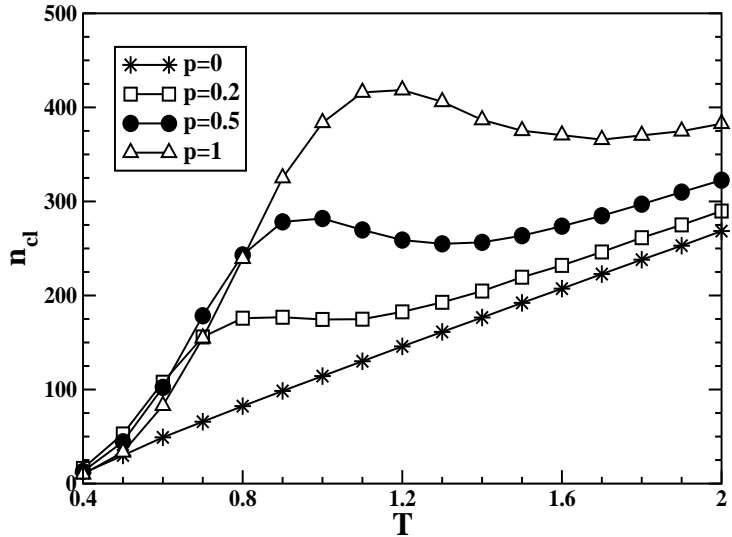}}}
\caption{Mean number of different clusters as a function of temperature, for a system of size $N=10^4$ 
and different values of the shortcut-adding probability, as indicated.}
\label{fig4}
\end{figure} 
 
Figure 4 shows the thermal dependence of the mean number of different clusters, $n_{cl}$. As in the previous Figure, 
results correspond to a system of size $N=10^4$ 
and different values of the shortcut-adding probability: $p=0, 0.2, 0.5,$ and $1$. 
The $p>0$ plots show broad local maxima taking place at intermediate temperatures. As with response functions such as 
the susceptibility and heat capacity (see Figure 5(b)-(c) below), the observed maxima could be indicative of   
peak fluctuations in the cluster structure due to thermally-driven, bulk order-disorder phase transitions. 
        
In order to gain further insight into these phase transitions, we will 
adopt an appropriate order parameter and focus on its thermal dependence, as well as on that of standard response functions. 
These results will be later extrapolated to the thermodynamic limit by means of finite-size scaling relations. 

The degree of order in a magnetic system can be naturally characterized by the ensemble-averaged magnetization per site-i.e.,
\begin{equation}
\langle M\rangle=\langle {\frac{1}{N}}\sum S_i\rangle \ .
\end{equation}
However, according to standard procedures to avoid spurious effects arising from the finite size of the simulated 
samples (see e.g. \cite{lan00,bin02}), it is more
appropriate to consider instead the ensemble-averaged absolute value of the magnetization per site, 
$\langle |M|\rangle$, as the order parameter. 

\begin{figure}[t]
\centerline{{\epsfxsize=6.2in \epsfysize=2.3in \epsfbox{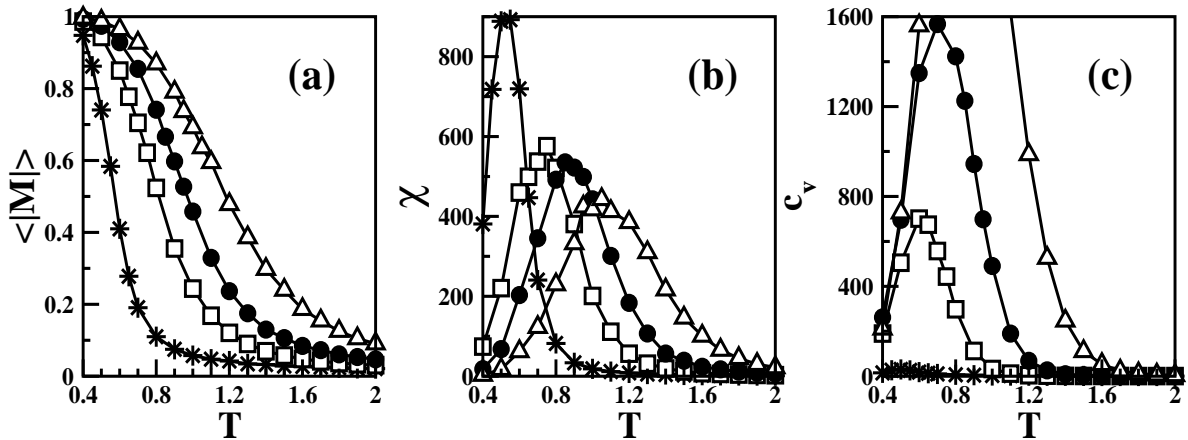}}}
\caption{Thermal dependence of: (a) the absolute magnetization; (b) the magnetic susceptibility; (c) the 
heat capacity. The system size is $N=10^4$. Different values for the shortcut-adding probability are shown: 
$p=0$ (stars), $p=0.2$ (open squares), $p=0.5$ (filled circles), and $p=1$ (open triangles).}
\label{fig5}
\end{figure}

Figure 5(a) shows plots of $\langle |M|\rangle$ as a function of $T$, for different values of $p$ and a fixed 
network size, $N=10^4$. As expected,  
the effect of increasing $p$ at a fixed temperature is that of increasing the net magnetization 
(and, hence, the order) of the system. Considering instead a 
fixed value of $p$, we see that, at low temperatures, the system grows ordered and the (absolute) magnetization
is close to unity, while at higher temperatures the disorder sets on and the net magnetization  
becomes significantly smaller. 
However, fluctuations due to the finite network size prevent the magnetization from 
becoming strictly zero above the critical temperature, and thus the transition between the low-temperature 
ordered phase and the high-temperature disordered one turns out to be smoothed out by finite-size effects. 

Figure 5(b) displays the thermal dependence of the order parameter fluctuations (or the analog of the 
``magnetic susceptibility" $\chi$), 
given by
\begin{equation}
\chi = \frac{N}{T}\left(\langle M^2\rangle-\langle|M|\rangle^2\right). 
\label{chimi}
\end{equation}
For equilibrium systems, the susceptibility is related to order parameter fluctuations by the 
fluctuation-dissipation theorem; however, the validity of a fluctuation-dissipation relation under far-from-equilibrium 
conditions is less evident. Keeping this caveat in mind, Eq.~(\ref{chimi}) proved useful to explore the critical 
behavior of several nonequilibrium spin models (see e.g. \cite{can01,sid98}), in which  
order-disorder phase transitions were indicated by the occurrence of peaks in the thermal dependence of $\chi$. 
In the same vein, Figure 5(c) shows the behavior of energy fluctuations (or the analog of the ``heat capacity" $c_v$), 
defined by
\begin{equation}
c_v =\frac{1}{NT^2}\left(\langle E^2\rangle-\langle E\rangle^2\right).
\label{heat}
\end{equation}

As with the thermal dependence of the order parameter shown in Figure 5(a), the order-disorder transitions
signaled by the peaks of the ``susceptibility" and ``heat capacity" become rounded and shifted. 
However, strictly speaking, 
Figure 5 is just showing evidence of pseudo-phase transitions, which might be precursors of true phase 
transitions taking place in the ($N\to\infty$) thermodynamic limit. 
Let us now proceed further and use standard finite-size 
scaling procedures to establish the phase diagram $T_c$ vs $p$ corresponding to the true phase 
transition in the thermodynamic limit.

According to the finite-size scaling theory, developed for the treatment of finite-size effects 
at criticality and under equilibrium conditions \cite{bar83, pri90},
the difference between the true critical temperature, $T_c$, and an effective pseudocritical 
one, $T_c^{eff}(N;p)$, is given by 
\begin{equation}
|T_c(p)-T_c^{eff}(N;p)|\propto N^{-1/2\nu}, 
\label{scaling}
\end{equation}
where $\nu$ is the exponent that characterizes the divergence of the correlation length 
at criticality. The effective pseudocritical temperature, $T_c^{eff}(N;p)$, is here defined as the 
value corresponding to $\langle |M|\rangle=0.5$ for a finite system of $N$ nodes and shortcut-adding 
probability $p$. 

\begin{figure}[t]
\centerline{{\epsfxsize=3.8in \epsfysize=2.8in \epsfbox{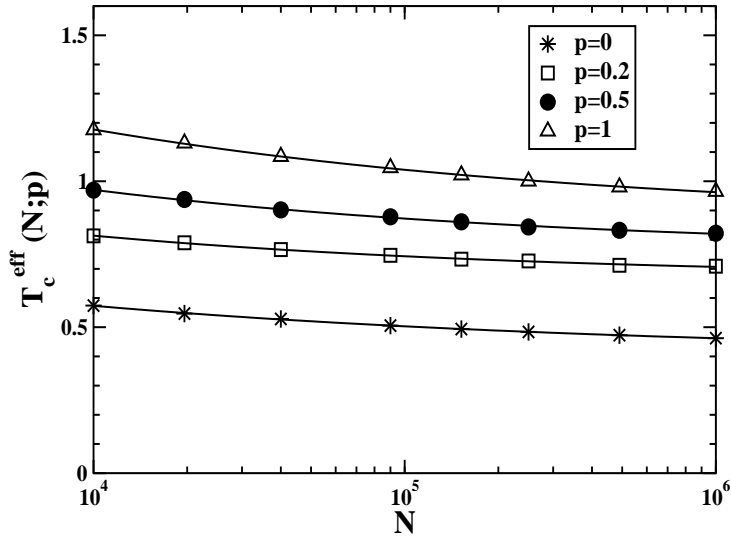}}}
\caption{Effective transition temperatures $T_c^{eff}(N;p)$ for $10^4\leq N\leq 10^6$ and different 
values of $p$, as indicated (symbols). Fits to the data using the finite-size scaling relation, 
Eq.~(\ref{scaling}), are also shown (solid lines). See more details in the text.}
\label{fig6}
\end{figure}

Figure 6 shows the effective transition temperatures (symbols) for small-world networks of different 
sizes and different shortcut-adding probabilities, as indicated. The solid lines show, for comparison, 
least-squares fits to the data, which where obtained using the finite-size scaling relation, Eq.~(\ref{scaling}).
The effective pseudocritical temperatures decrease monotonically with the system size 
and are indeed observed to extrapolate to finite critical temperatures in the thermodynamic limit. 
This behavior is qualitatively similar to previous studies on the MEM growing on 1D small-world networks 
with $p>0$ \cite{can06}. However, the regular-lattice limit showed that the MEM was noncritical in 1D, 
while here, in 2D, the system also exhibits critical behavior in the absence of shortcuts ($p=0$).   

\begin{figure}[t]
\centerline{{\epsfxsize=3.8in \epsfysize=2.8in \epsfbox{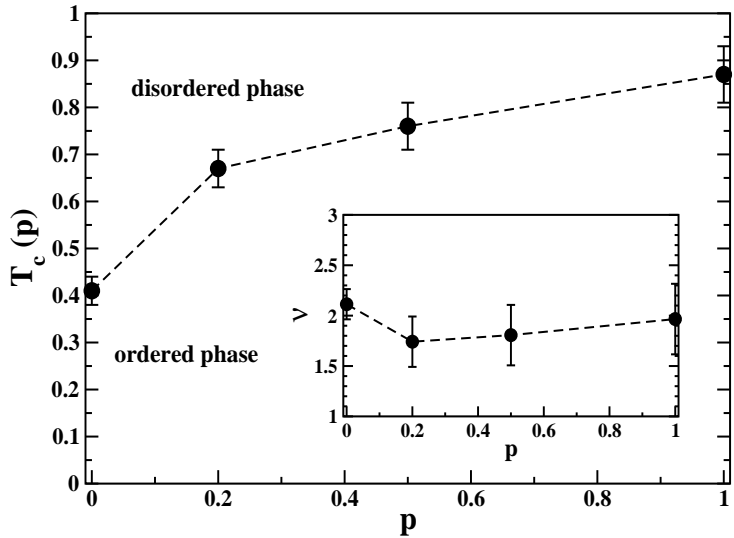}}}
\caption{Critical temperature as a function of the shortcut-adding probability $p$. The inset shows 
the correlation length critical exponent $\nu$. The dashed lines are guides to the eye.}
\label{fig7}
\end{figure}

Figure 7 shows the critical temperatures obtained for different values of the shortcut-adding 
probability $p$, which separate the ordered phase from the disordered one. 
The error bars reflect statistical 
errors, which were determined from the fitting procedure.  
Moreover, the inset shows the correlation length critical exponent $\nu$, which is also determined from 
the finite-size scaling relation, Eq.~(\ref{scaling}). 
Within error bars, the critical exponent $\nu\approx 2$
is not found to depend on the shortcut density, in agreement with results from previous investigations on 
spin systems defined on small-world networks \cite{her02,med03}. 

\section{Conclusions}
Motivated by irreversible opinion spreading phenomena, we investigated the magnetic Eden model (MEM) growing 
on small-world networks generated from 2D regular lattices. 
In this nonequilibrium model, the opinion or decision of an individual 
is affected by those of their acquaintances, but opinion changes (analogous to spin flips in an Ising-like model) 
do not occur. 

The model studied in this work could be realistically applied to sociological scenarios in which individuals are subject 
to highly polarized, short term, binary choice situations. Given these conditions opinions are not expected to fluctuate 
and ``thermalize". One example is a binary voting scenario, such as a ballotage or referendum, where the
validity of this model could be empirically tested using time-resolved data from polls and surveys.

Dynamical processes of clustering and formation of opinion domains were first discussed, on a 
qualitative basis, by means of snapshot configurations, and were later addressed quantitatively by measurements of cluster size 
probability distributions, the size of the largest cluster, and the number of different clusters. Results showed
a strong dependence on both the ``temperature" (i.e. the strength of an individual's free will relative to the homophilic 
ties to their social neighbors) as well as with the number of small-world shortcut connections. 
In agreement with related investigations on spin systems on small-world networks, we observed the existence of 
thermally-driven, order-disorder transitions, in which the ordering effects increase monotonically with the number of 
shortcuts. However, unlike the MEM growing on 1D small-world networks, the 2D system 
is critical also in the absence of shortcuts. By means of standard finite-size extrapolation procedures, we obtained 
the order-disorder phase diagram extrapolated to $N\to\infty$, as well as the correlation length critical 
exponent. 

The magnetic Eden model, being different in nature from related equilibrium spin systems such as the Ising model, 
can certainly provide valuable, complementary insight into dynamical and critical aspects of the spreading of opinions 
in a society. Hopefully, the present findings will thus contribute to the growing interdisciplinary efforts in 
the fields of sociophysics, complex networks, and nonequilibrium statistical physics, and stimulate further work. 

\section*{Acknowledgments} 
The author thanks K. Mazzitello for providing an algorithm for cluster identification. 
This work was supported by the James S. McDonnell Foundation and the National 
Science Foundation ITR DMR-0426737 and CNS-0540348 within the DDDAS program.

\end{document}